\documentclass[12pt]{article}
\usepackage[a4paper,margin=2cm]{geometry}
\usepackage{cmap}
\usepackage{concmath}
\usepackage{amsmath,amssymb}
\usepackage{cite,icomma,indentfirst}
\usepackage{graphicx,overpic,wrapfig}
\usepackage[labelsep=period,labelfont=bf]{caption}
\usepackage[utf8]{inputenc}
\usepackage[T1,T2A]{fontenc}
\usepackage[english,russian]{babel}

\newcommand{\code}[1]{``\texttt{#1}''}
\newcommand{\vb}[1]{\boldsymbol{#1}}
\newcommand{\hm}[1]{#1\nobreak\discretionary{}{\hbox{\ensuremath{#1}}}{}}
\newcommand{\w}{\linewidth}
\newcommand{\T}{^\mathrm{T}}
\newcommand{\Ci}[2]{\mathbf{I}_{#1}(#2)}
\newcommand{\pdif}{\partial}
\newcommand{\idx}[1]{1, 2, \ldots, {#1}}
\newcommand{\idyi}{0, 1, \ldots}
\DeclareMathOperator{\lb}{lb}
\DeclareMathOperator{\dif}{d\!}

\usepackage[pdftex,unicode]{hyperref}
\hypersetup{unicode=true,
pdfauthor={P.V. Moskalev},
pdfsubject={Estimates of threshold and strength of percolation clusters on square lattices with (1,d)-neighborhood},
pdftitle={Moskalev P.V. Estimates of threshold and strength of percolation clusters on square lattices with (1,d)-neighborhood},
pdfkeywords={site percolation, square lattice, non-metric Minkowski distance, Moore neighborhood, percolation threshold, strength of infinite clusters, R programming language, SPSL package}}

\begin{document}

\begin{titlepage}

\English
\begin{flushleft}\bfseries
UDC 519.676
\end{flushleft}
\begin{flushleft}\bfseries\Large
Estimates of threshold and strength of percolation clusters on square lattices with $\boldsymbol{(1, d)}$-neighborhood
\end{flushleft}
\begin{flushleft}
\textbf{P.V. Moskalev}\\[1ex]
Voronezh State Agricultural University (moskalefff@gmail.com)
\end{flushleft}

\noindent\textbf{Abstract.} In this paper we consider statistical estimates of threshold and strength of percolation clusters on square lattices.
The percolation threshold $p_c$ and the strength of percolation clusters $P_\infty$ for a square lattice with $(1, d)$-neighborhood depends not only on the lattice dimension, but also on the Minkowski exponent $d$.
To estimate the strength of percolation clusters $P_\infty$ proposed a new method of averaging the relative frequencies of the target subset of lattice sites.
The implementation of this method is based on the SPSL package, released under GNU GPL-3 using the free programming language R.
\smallskip

\medskip

\Russian
\begin{flushleft}\bfseries
УДК 519.676
\end{flushleft}
\begin{flushleft}\bfseries\Large
Оценки порога и мощности перколяционных кластеров на квадратных решётках с $\boldsymbol{(1, \pi)}$-окрестностью
\end{flushleft}
\begin{flushleft}
\textbf{П.В. Москалев}\\[1ex]
Воронежский государственный аграрный университет (moskalefff@gmail.com)
\end{flushleft}

\noindent\textbf{Аннотация.}
В работе рассматриваются статистические оценки порога и мощности перколяционных кластеров на квадратных решётках.
Порог перколяции $p_c$ и мощность перколяционных кластеров $P_\infty$ на квадратной решётке с $(1, \pi)$-окрестностью зависят не только от размерности решётки, но от показателя Минковского $\pi$.
Для оценки мощности перколяционных кластеров $P_\infty$ предложен новый метод, основанный на усреднении относительных частот целевого подмножества узлов решётки.
Реализация предложенного метода основана на библиотеке SPSL, выпущенной под лицензией GNU GPL-3 с использованием свободного языка программирования R.
\smallskip

\noindent\textbf{Ключевые слова:} перколяция узлов, квадратная решётка, неметрическое расстояние Минковского, окрестность Мура, порог перколяции, мощность перколяционного кластера, язык программирования R, библиотека SPSL


\normalsize

\end{titlepage}

\section{Общие определения}

Простейшая модель изотропной перколяции узлов строится с помощью взвешенного однородного графа (перколяционной решётки), достижимость произвольного узла которой задается неравенством $u_i<p$, где $u_i\sim \mathbf{U}(0, 1)$~--- определяющий достижимость узла весовой коэффициент; $p\in [0, 1]$~--- определяющая вероятность протекания доля достижимых узлов.
Одним из результатов решения задачи о построении связанных подмножеств (кластеров узлов) для выборочной совокупности ограниченных решёток является оценка относительной частоты $w$ возникновения пути, связывающего стартовое и целевое подмножества узлов решётки, при заданных параметрах решётки.
Подмножество узлов, связывающих противолежащие стороны ограниченной перколяционной решётки в теории протекания называется стягивающим кластером.
Относительная частота появления стягивающего кластера используется для оценки порога протекания $p_c$, соответствующего вероятности возникновения неограниченного кластера.

Помимо порога протекания $p_c$ существенное значение для прикладных исследований могут иметь и другие характеристики кластеров, например, массовая фрактальная размерность $d_M$ мощность $P_\infty$ \cite{tarasevich.2002.percolation}.
Как показывают наши исследования \cite{moskaleff.info.2011.ssi20.pc.pi, moskaleff.info.2013.ssi20.pinf} все эти характеристики зависят не только от типа и топологической размерности перколяционной решётки, но и от формы $(1, \pi)$-окрестности её узлов.

Для единичной окрестности узла на квадратной решётке известны две дискретные формы: минимальная $(1, 0)$-окрестность фон Неймана и максимальная $(1, \infty)$-окрест\-ность Мура.
$(1, 0)$-окрестность фон Неймана образуется как подмножество узлов решётки, только одна из координат которых отличается от одноимённой координаты выделенного узла на единицу, а в $(1, \infty)$-окрестность Мура входят узлы, хотя бы одна из координат которых отличается от одноимённой координаты выделенного узла на единицу.
Из сделанных определений следует, что $(1, 0)$-окрестность фон Неймана является подмножеством $(1, \infty)$-окрестности Мура.
Более того в работах \cite{moskaleff.mce.2013.ssTNd.perc, moskaleff.crm.2013.ssTNd} было показано, что на базе классической $(1, \infty)$-окрестности Мура можно построить более общий вариант $(1, \pi)$-окрестности Мура.

Из теории множеств известно \cite{alexandrov.1977.topology}, что ключевое влияние на структуру множества оказывает функция метрики, определяющая расстояния и формирующая $\varepsilon$-окрестность некоторой точки $b$.
Одним из достаточно общих способов определения окрестности произвольной точки $U_{\varepsilon, \pi}(b)$ является использование функции неметрического расстояния Минковского $\rho_\pi(a, b)$:
\begin{gather}\label{eq:mink_metric}
U_{\varepsilon, \pi}(b) = \{a : \rho_\pi(a, b)\leqslant\varepsilon\}, \quad
\rho_\pi(a, b) = \biggl(\sum_{i=1}^k |a_i - b_i|^\pi\biggr)^{1/\pi},
\end{gather}
где $\pi\geqslant 0$~--- показатель неметрического расстояния Минковского (для краткости далее по тексту именуемый просто показателем Минковского); $a(a_1$, $a_2$, \ldots, $a_k)$, $b(b_1$, $b_2$, \ldots, $b_k)$~--- координаты точек $a$ и $b$.
Применение термина <<неметрическое расстояние>> обусловлено тем, что строгое определение метрики накладывает на функцию \eqref{eq:mink_metric} следующие ограничения:
а)~$\rho_\pi(a, b)\hm=0 \Leftrightarrow a=b$;
б)~$\rho_\pi(a, b)=\rho_\pi(b, a)$;
в)~$\rho_\pi(a, b)\hm\leqslant \rho_\pi(a, c)+\rho_\pi(c, b)$.
Для неметрического расстояния Минковского все три ограничения выполняются лишь при $\pi\geqslant 1$, а на интервале $0\leqslant\pi<1$ знак в третьем неравенстве (неравенстве треугольника) меняется на противоположный $\rho_\pi(a, b)\hm> \rho_\pi(a, c)+\rho_\pi(c, b)$.
В наших задачах функция неметрического расстояния $\rho_\pi(a, b)$ определяет лишь меру удалённости точек $a$ и $b$ вдоль проходящей через них прямой и используется в \eqref{eq:mink_metric} для определения окрестности $b$ с соответствующим показателем Минковского $\pi$.

Тогда, для определения достижимости узлов, образующих $(1, \pi)$-окрестность Мура, с учётом меры их удалённости от текущего узла $\rho_\pi$ вероятностное неравенство обобщённой перколяционной модели примет вид $u_i<\frac{p}{\rho_\pi}$.

\section[Оценка порога перколяции c (1,п)-окрестностью]{Оценка порога перколяции c $\boldsymbol{(1, \pi)}$-окрестностью}

Рассмотрим задачу статистической оценки значения порога перколяции по выборочной совокупности реализаций кластеров узлов на двумерных квадратных решётках. 
При протекании в заданном направлении конечномерная оценка порога перколяции $p_c$ соответствует доле достижимых узлов $p$ ограниченной решётки, при которой частота появления стягивающего кластера $w$ достигает медианного значения $w(p=p_c)=\frac{1}{2}$.
Как было отмечено в докладе \cite{moskaleff.info.2011.ssi20.pc.pi} расширение $(1, 0)$-окрест\-ности фон Неймана до $(1, \pi)$-окрестности Мура приводит к тому, что частота появления стягивающих кластеров $w$ становится зависимой не только от доли достижимых узлов $p$ и размера квадратной решётки $L$, но и от показателя Минковского $\pi$.

\begin{figure}[bh]\centering\small
\begin{overpic}{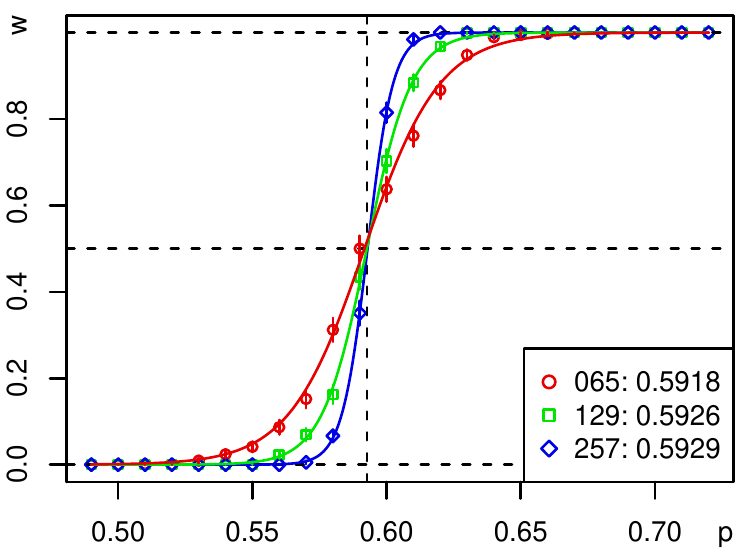}
\put(89,30){а)}
\end{overpic}\quad
\begin{overpic}{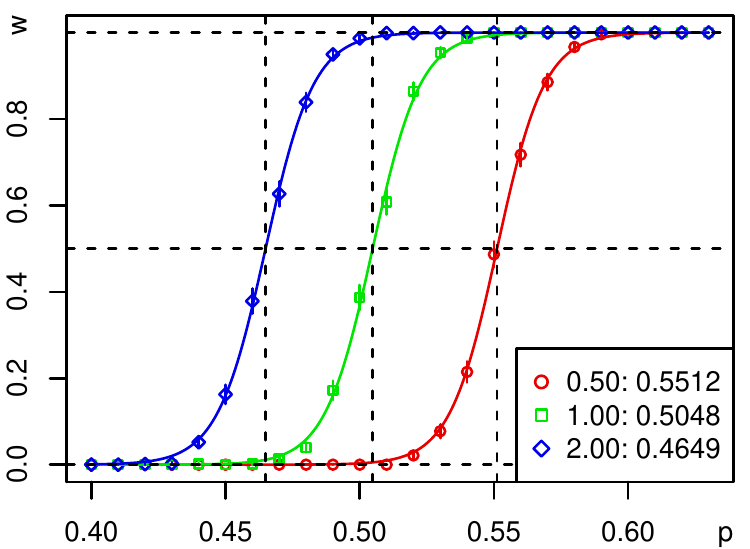}
\put(89,30){б)}
\end{overpic}
\caption{\label{pic:w_plpi}
Относительные частоты кластеров $w(p|L, \pi)$, стягивающих решётку от $y=-\frac{L}{2}$ до $y=\frac{L}{2}$ при: а)~$\pi=0$ и $L=65$, $129$, $257$ узлов; б)~$\pi=\frac{1}{2}$, $1$, $2$ и $L=129$ узлов}
\end{figure}

На рис.\,\ref{pic:w_plpi} приведены результаты статистического моделирования перколяции узлов в направлении от нижней $y=-\frac{L}{2}$ до верхней $y=\frac{L}{2}$ границ двумерных квадратных решёток с $(1, \pi)$-окрестностью при различных значениях размера решётки $L$ и показателя Минковского $\pi$.
На рис.\,\ref{pic:w_plpi} (а) показаны зависимости относительных частот возникновения стягивающих кластеров $w$ на решётках размерами $L=65$, $129$ и $257$ узлов с $(1, 0)$-окрестностью фон Неймана от доли достижимых узлов $p$.
Символами ``\scalebox{.8}{$\bigcirc$}'' показаны зависимости $w(p|L, \pi)$ для решётки размером $L_1\hm=65$ узлов, а символами ``$\square$'' и ``\raisebox{-.4ex}{\rotatebox{45}{$\square$}}''~--- зависимости $w(p|L, \pi)$ для решёток с размерами $L_2=129$ и $L_3=257$ узлов соответственно.

На рис.\,\ref{pic:w_plpi} (б) показаны зависимости относительных частот возникновения стягивающих кластеров $w$ на решётках размера $L=129$ узлов с $(1, \frac{1}{2})$-, $(1, 1)$- и $(1, 2)$-окрестностью от доли достижимых узлов $p$.
Символами ``\scalebox{.8}{$\bigcirc$}'' на рис.\,\ref{pic:w_plpi} (б) показаны зависимости $w(p|L, \pi)$ для решётки с $(1, \frac{1}{2})$-окрестностью, а символами ``$\square$'' и ``\raisebox{-.4ex}{\rotatebox{45}{$\square$}}''~--- зависимости $w(p|L, \pi)$ для решёток с $(1, 1)$- и $(1, 2)$-окрестностями соответственно.

Объёмы выборок, используемых для расчёта относительной частоты возникновения стягивающих кластеров $w$, для каждой тройки значений $(p|L, \pi)$ на рис.\,1 составляет $n=750$ реализаций.
Вертикальные отрезки соответствуют радиусам $0,95$-доверительных интервалов для наблюдаемых значений $w$.
При построении интервальной оценки для относительной частоты $w$ учитывается, что при фиксированных параметрах $p$, $L$ и $\pi$ вероятность возникновения стягивающего кластера будет постоянна.
Тогда при независимых по реализациям распределениях псевдослучайных весовых коэффициентов $u_{xy}$ оцениваемая случайная величина $W$ будет подчиняться биномиальному распределению, а её доверительный интервал примет вид:
\begin{gather}\label{eq:Ci_w}
\Ci{0,95}{W}\hm=\Bigl(w\pm t_{0,95}\sqrt{\tfrac{1}{n}w(1-w)}\Bigr),
\end{gather}
где $w$~--- относительная частота стягивающего кластера; $n$~--- объём выборки; $t_{0,95}\hm\approx 1,6449$~--- $0,95$-квантиль стандартного нормального распределения.

По рис.\,\ref{pic:w_plpi} (а) нетрудно заметить, что при протекании от нижней $y=-\frac{L}{2}$ до верхней $y=\frac{L}{2}$ границ решётки интервал оси абсцисс $p$, на котором функция $w(p|L, \pi)$ отлична от нуля и единицы, сужается по мере роста $L$, стягиваясь при $L\hm\to\infty$ к известному значению порога протекания для двумерной квадратной решётки $p_c=0,592746\ldots$, который показан вертикальной штриховой линией:
\begin{gather}\label{eq:lim_w(p|L)}
\lim_{L\to\infty} w(p|L) =
\begin{cases}
0,           & p < p_c;\\
\frac{1}{2}, & p = p_c;\\
1,           & p > p_c.
\end{cases}
\end{gather}

Кривые, аппроксимирующие расчётные точки, соответствуют графикам нормированной логистической функции:
\begin{gather}\label{eq:logit}
f(\vb p|\vb\alpha) = \frac{1}{1+e^{-\frac{\vb p-\alpha_1}{\alpha_2}}},
\end{gather}
где $\vb\alpha=(\alpha_1, \alpha_2)$~--- вектор параметров, оцениваемый с помощью нелинейной регрессионной модели $\vb w = f(\vb p|\vb\alpha) + \vb e$ c евклидовой нормой невязки $\vb e\hm=|\vb w - f(\vb p|\vb\alpha)|^2\hm\to\min$, минимизируемой методом Ньютона--Гаусса \cite{bates.1988.nonlinear.regression}.
После линеаризации модели с помощью разложения функции $f(\vb p|\vb\alpha)$ по компонентам вектора $\vb\alpha$ в ряд Тейлора искомая оценка параметров на $k+1$ итерации примет вид:
$\vb\alpha_{k+1} = (\vb J_k\T \vb J_k)^{-1} (\vb J_k\T \vb w)$,
где $\vb J_k$~--- матрица Якоби, состоящая из значений частных производных аппроксимирующей функции $J_{ijk}=\frac{\pdif f(p_i|\vb\alpha_k)}{\pdif\alpha_{jk}}$ при $i=\idx{n}$, $j=1, 2$, $k=\idyi$,
здесь $i, j$~--- индексы элементов выборки и вектора параметров на $k$ итерации; $\vb p, \vb w$~--- векторы выборочных данных.
В результате при использовании аппроксимации вида \eqref{eq:logit} сдвиговая компонента вектора $\vb\alpha$ будет соответствовать искомой оценке порога протекания $\alpha_1\approx p_c$.
Заметим, что при неограниченном возрастании масштабной компоненты $\alpha_2\to\infty$ предел аппроксимирующей функции \eqref{eq:logit} будет совпадать с медианным значением относительной частоты \eqref{eq:lim_w(p|L)}:
$\lim_{\alpha_2\to\infty} f(\vb p|\vb\alpha) = \frac{1}{2}$.

В приведённых выше расчётах были использованы функции \code{ssi20()} и \code{ssi2d()} из состава библиотеки SPSL \cite{moskalev.cran.2012.spsl}, выпущенной автором под лицензией GNU GPL-3 и доступной для свободной загрузки через систему репозиториев CRAN.

Теперь рассмотрим оценку термодинамического предела порога перколяции $p_c$ с помощью скейлинга на ограниченных двумерных квадратных решётках для описанного в \cite{moskaleff.mce.2013.ssTNd.perc, moskaleff.crm.2013.ssTNd} обобщения модели изотропной перколяции узлов с $(1, \pi)$-окрестностью.
Как уже было отмечено выше расширение $(1, 0)$-окрестности фон Неймана до $(1, \pi)$-окрестности Мура приводит к тому, что  определяемый по частоте появления стягивающих кластеров $w$ порог перколяции $p_c$ становится зависимым не только от размера квадратной решётки $L$, но и от показателя Минковского $\pi$.

Термодинамическим пределом порога перколяции называют оценку $p_c(L|\pi)$, возникающую при неограниченном увеличении размера перколяционной решётки $L\to\infty$.
В классических работах по скейлинговой теории \cite{stauffer.1979.scaling} для оценки термодинамического предела порога перколяции используется известное соотношение, которое в нашем случае можно записать:
\begin{gather*}
\bigl|\,p_c(L|\pi) - p_c(\pi)\,\bigr| \propto L^{-1/\nu}, 
\end{gather*}
где $\nu$~--- универсальный скейлинговый показатель, для плоских решёток равный $\nu=\frac{4}{3}$. 
Однако, как было отмечено в статье \cite{ziff.2001.threshold}, более быструю сходимость конечномерных оценок $p_c(L|\pi)$ к своим термодинамическим пределам $p_c(\pi)$ обеспечивает модифицированное соотношение:
\begin{gather}\label{eq:scaling}
\bigl|\,p_c(L|\pi) - p_c(\pi)\,\bigr| \propto L^{-2-1/\nu}.
\end{gather}

\begin{figure}[tbh]\centering
\begin{overpic}[width=.32\w]{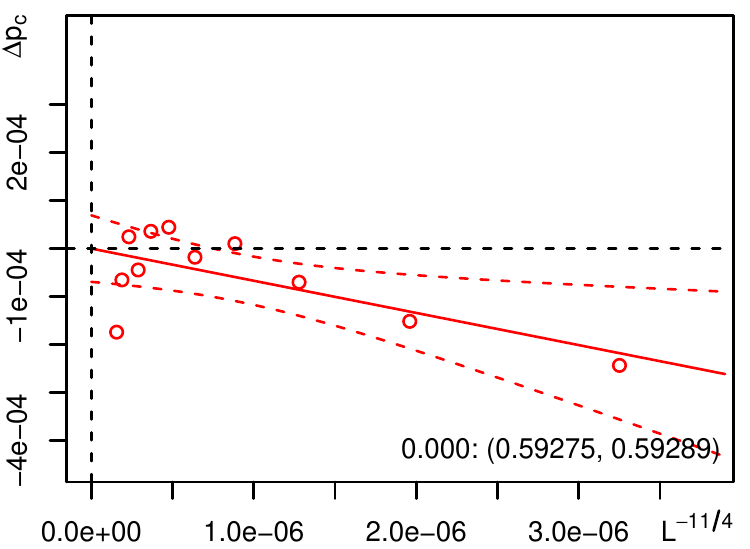}
\put(87,60){а)}
\end{overpic}
\begin{overpic}[width=.32\w]{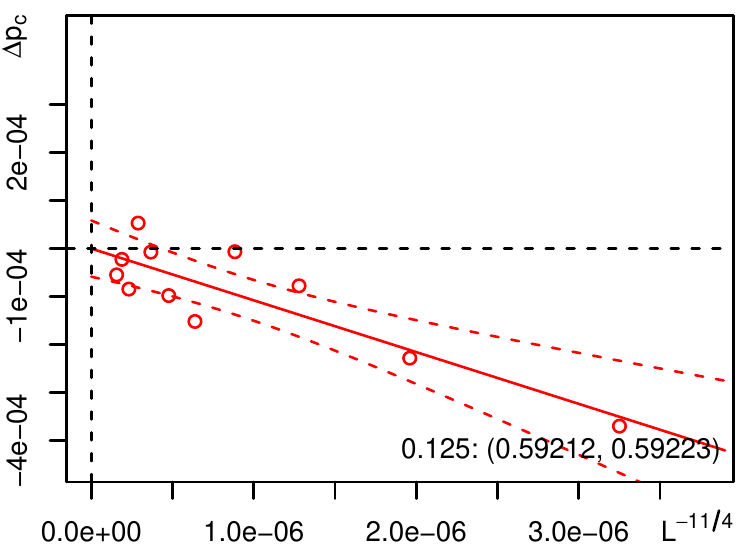}
\put(87,60){б)}
\end{overpic}
\begin{overpic}[width=.32\w]{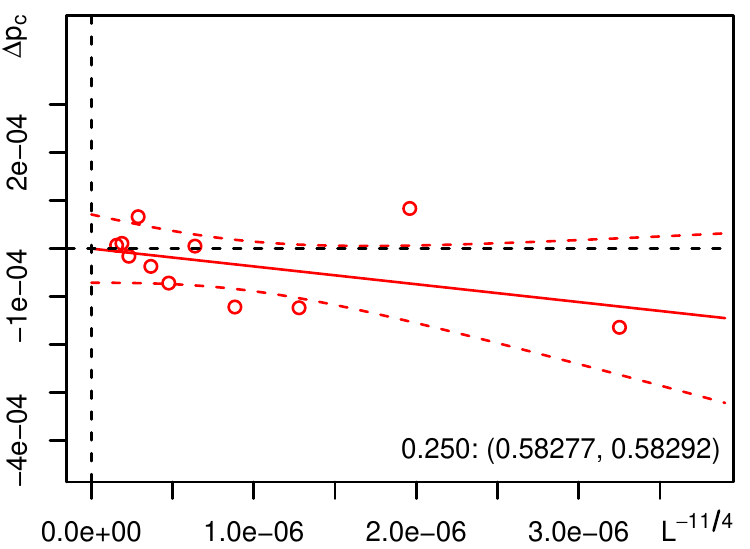}
\put(87,60){в)}
\end{overpic}\\
\begin{overpic}[width=.32\w]{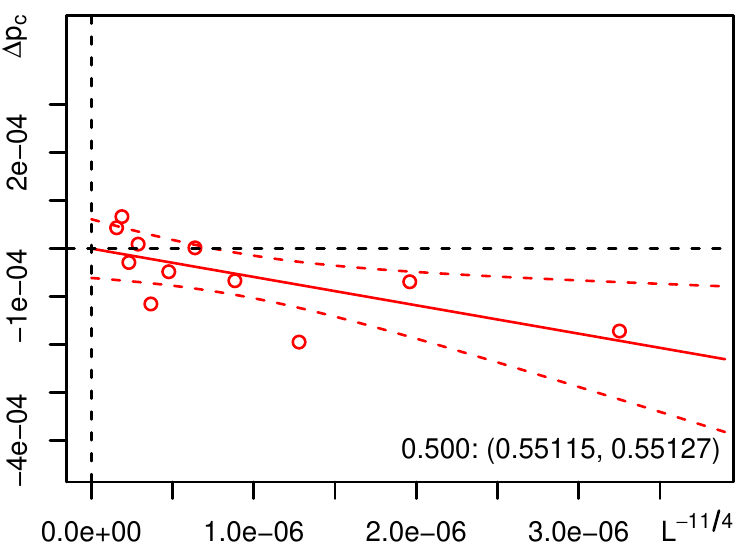}
\put(87,60){г)}
\end{overpic}
\begin{overpic}[width=.32\w]{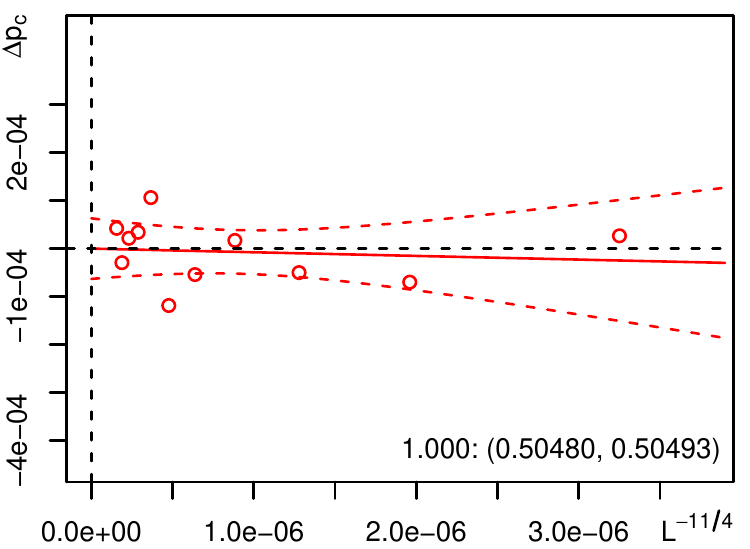}
\put(87,60){д)}
\end{overpic}
\begin{overpic}[width=.32\w]{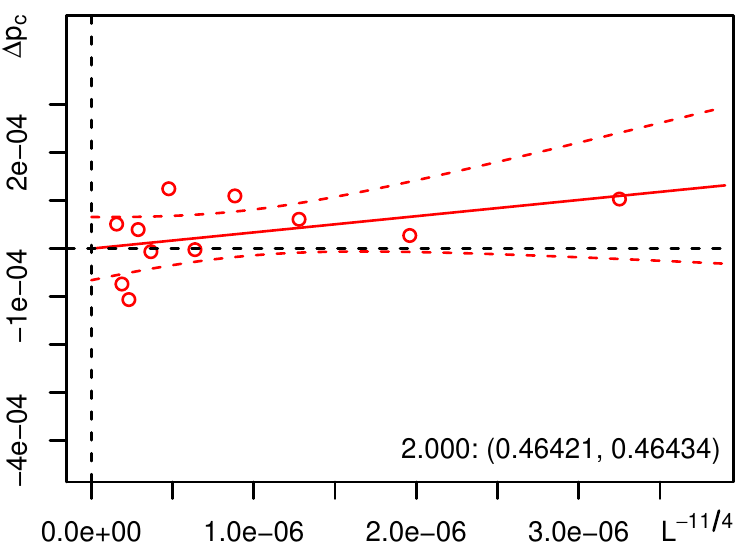}
\put(87,60){е)}
\end{overpic}\\
\begin{overpic}[width=.32\w]{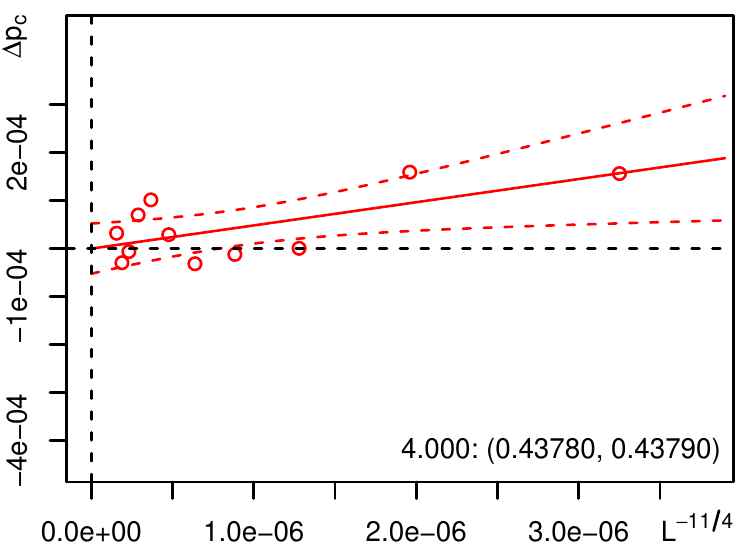}
\put(87,60){ж)}
\end{overpic}
\begin{overpic}[width=.32\w]{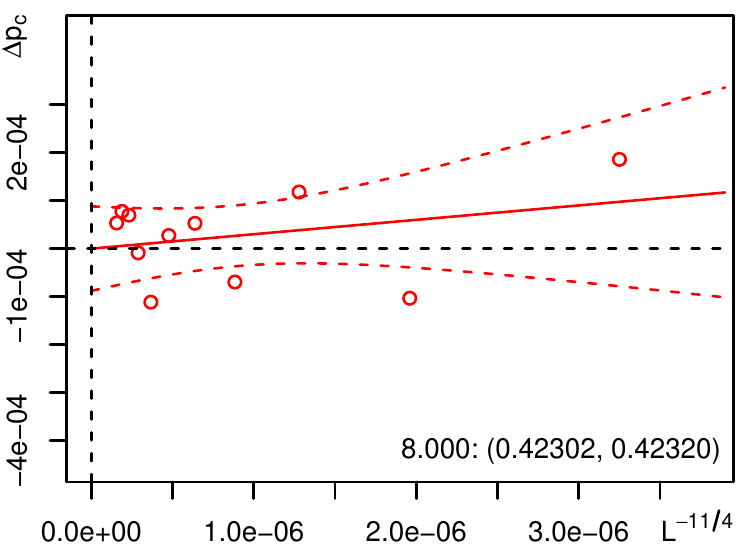}
\put(87,60){з)}
\end{overpic}
\begin{overpic}[width=.32\w]{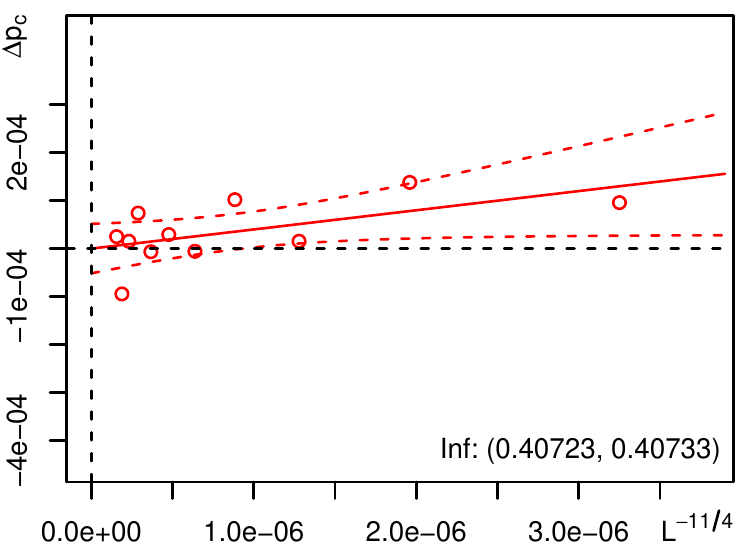}
\put(87,60){и)}
\end{overpic}
\caption{\label{pic:scaling}
Скейлинговые оценки $\Delta p_c(L^{-11/4}|\pi)$ на решётках размером $L=101$, 151, \ldots, 451 узлов при значениях показателя Минковского: а)~$\pi=0$; б)~$\pi=\frac{1}{8}$; в)~$\pi=\frac{1}{4}$; г)~$\pi=\frac{1}{2}$; д)~$\pi=1$; е)~$\pi=2$; ж)~$\pi=4$; з)~$\pi=8$; и)~$\pi\to\infty$}
\end{figure}

Примеры интервальных скейлинговых оценок $p_c(\pi)$ по выборкам объёмом $n=12000$ реализаций на двумерных квадратных решётках с размерами $L\hm=101$, $151$, \ldots, $451$ узлов с $(1, \pi)$-окрестностью при $\pi=0$, $\frac{1}{8}$, $\frac{1}{4}$, $\frac{1}{2}$, 1, 2, 4, $8$, $\infty$ показаны на рис.\,\ref{pic:scaling} (а-и). 
В соответствии с \eqref{eq:scaling} по оси абсцисс откладывались значения степенной функции размера решётки $L^{-11/4}$, а по оси ординат~--- отклонений $\Delta p_c(L^{-11/4}|\pi)$ конечномерных оценок порогов перколяции $p_c(L^{-11/4}|\pi)$ от своих термодинамических пределов $p_c(\pi)$.

Символами ``\scalebox{.8}{$\bigcirc$}'' на рис.\,\ref{pic:scaling} показаны расчётные точки, сплошными прямыми линиями~--- линии регрессии, а штриховыми гиперболическими линиями~--- 0,95-доверительные интервалы для моделей нормальной линейной регрессии $\Delta p_c$ на $L^{-11/4}$. 
Соответствующие интервальные оценки термодинамического предела порога перколяции $p_c(\pi)$ при различных значениях показателя Минковского $\pi$ указаны в левом нижнем углу каждого графика на рис.\,\ref{pic:scaling}:
а)~$\Ci{0,95}{p_c(0)}=(0,59275$; $0,59289)$;
б)~$\Ci{0,95}{p_c(\frac{1}{8})}=(0,59212$; $0,59223)$;
в)~$\Ci{0,95}{p_c(\frac{1}{4})}=(0,58277$; $0,58292)$;
г)~$\Ci{0,95}{p_c(\frac{1}{2})}=(0,55115$; $0,55127)$;
д)~$\Ci{0,95}{p_c(1)}\hm=(0,50480$; $0,50493)$;
е)~$\Ci{0,95}{p_c(2)}=(0,46421$; $0,46434)$;
ж)~$\Ci{0,95}{p_c(4)}=(0,43780$; $0,43790)$;
з)~$\Ci{0,95}{p_c(8)}=(0,42302$; $0,42320)$;
и)~$\Ci{0,95}{p_c(\infty)}=(0,40723$; $0,40733)$.

Нетрудно заметить, что при неметрических показателях Минковского $\pi<1$ значения коэффициентов регрессии отрицательны, при метрических показателях $\pi>1$ коэффициенты регрессии положительны, а в манхеттенской метрике при $\pi=1$ значение коэффициента регрессии близко к нулю.

\begin{wrapfigure}{R}{.4\w}\centering
\includegraphics[width=.95\w]{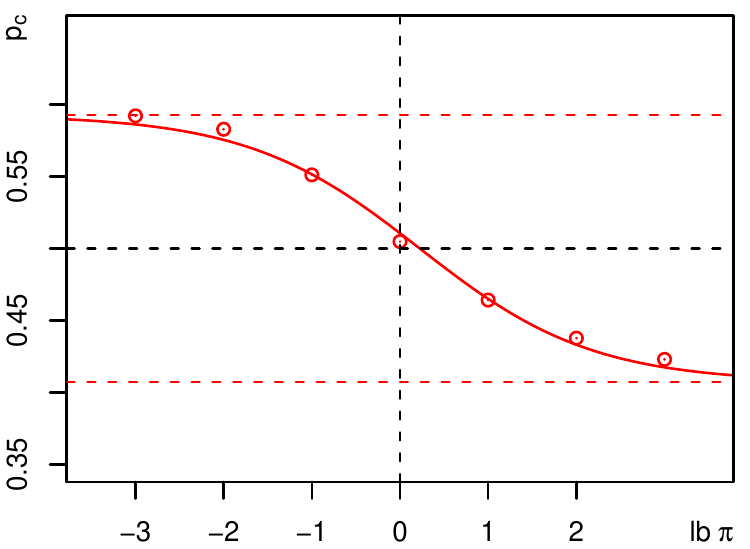}
\caption{\label{pic:pc_lbpi}
Скейлинговые оценки порога перколяции узлов $p_c(\lb\pi)$ на двумерной квадратной решётке с $(1, \pi)$-окрестностью}
\end{wrapfigure}

На рис.\,\ref{pic:pc_lbpi} показана зависимость скейлинговой оценки порога протекания от двоичного логарифма показателя Минковского $p_c(\lb\pi)$ в задачах узлов для двумерной квадратной решётки.
Показатель Минковского изменялся в пределах $\pi\in [0, \infty)$.
Полученные оценки показывают, что зависимость $p_c(\lb\pi)$ имеет монотонный характер, ограниченный по ординате двумя односторонними асимптотами на уровнях $p_c(\infty)\hm\approx 0,40728$ и $p_c(-\infty)\hm\approx 0,59282$.
Заметим, что последняя оценка удовлетворительно согласуется с известным из литературных источников \cite{lee.2008.threshold} значением порога протекания в классической задаче узлов на двумерной квадратной решётке $p_c\hm\approx 0,5927464(5)$.

Нетрудно проверить, что представленная на рис.\,\ref{pic:pc_lbpi} зависимость $p_c(\lb\pi)$ в первом приближении может быть аппроксимирована логистической функцией вида:
\begin{gather}\label{eq:pc(lb_pi)}
p_c(\lb\pi|\vb \alpha) = p_c(-\infty) -
\frac{p_c(-\infty)-p_c(\infty)}{1 + \exp(-\frac{\lb\pi-\alpha_1}{\alpha_2})},
\end{gather}
где 
$p_c(-\infty) = \lim\limits_{\pi\to 0+} p_c(\lb\pi)$,
$p_c(\infty) = \lim\limits_{\pi\to\infty} p_c(\lb\pi)$~--- скейлинговые оценки порога перколяции $p_c$, достигаемые при предельных значениях показателя Минковского $\pi$; 
$\vb\alpha\hm=(\alpha_1, \alpha_2)$~--- вектор параметров аппроксимации.

Из выражения \eqref{eq:pc(lb_pi)} следует, что зависимость $p_c(\lb\pi)$ в первом приближении должна быть симметрична некоторой относительно точки $(\lb\pi_0, p_{c0})$.
Для перколяции узлов на двумерной квадратной решётке с $(1, \pi)$-окрестностью  точка симметрии зависимости \eqref{eq:pc(lb_pi)} лежит в области значений $\pi$, близких к манхеттенской метрике, с оценкой порога протекания $p_c(\pi=1)\hm\approx 0,50487$, которая оказывается достаточно близкой к известному значению порога протекания \cite{stauffer.1979.scaling} в задаче связей на той же решётке $p'_c=\frac{1}{2}$.

\section[Оценка мощности перколяционных кластеров с (1,п)-окрестностью]{Оценка мощности перколяционных кластеров с $\boldsymbol{(1, \pi)}$-окрестностью}

Формальное определение порога перколяции $p_c$ основывается на использовании функции мощности перколяционного кластера $P_\infty(p)$, которая соответствует вероятности того, что случайно выбранный узел решётки будет принадлежать перколяционному кластеру.
На неограниченной решётке мощность перколяционного кластера $P_\infty$ будет равна нулю в докритической и строго больше нуля в сверхкритической области.
Тогда порог перколяции $p_c$ определяется как точная верхняя грань множества значений $p$, для которых мощность перколяционных кластеров будет равна нулю:
\begin{equation}\label{eq:pc_inf}
p_c=\sup\{p: P_\infty(p)=0\},
\quad\text{где}\ \
P_\infty(p) \begin{cases}
= 0, & p \leqslant p_c;\\
> 0, & p > p_c.
\end{cases}
\end{equation}

Как было отмечено в работе \cite{moskaleff.info.2013.ssi20.pinf} из определения \eqref{eq:pc_inf} следует, что при возрастании доли достижимых узлов $p\to 1-$ левосторонние пределы как для мощности перколяционного кластера, так и для её первой производной будут совпадать с пределом своего аргумента:
\begin{equation}\label{eq:lim_P_infty}
\lim\limits_{p\to 1-} P_\infty(p) =
\lim\limits_{p\to 1-} \frac{\dif P_\infty(p)}{\dif p} = 1.
\end{equation}

В скейлинговой теории \cite{stauffer.1979.scaling} считается, что мощность перколяционного кластера $P_\infty$ в окрестности критической точки при $p\to p_c$ пропорциональна расстоянию до неё с соответствующим показателем:
\begin{equation}\label{eq:P_infty_scaling}
P_\infty(p|p_c, \beta)\propto |p-p_c|^\beta,
\end{equation}
где $\beta$~--- зависящий от размерности пространства критический показатель, для двумерных решёток равный $\beta=\frac{5}{38}$.
Сохраняя общую форму зависимости \eqref{eq:P_infty_scaling} предположим, что мощность кластера в сверхкритической области при $p>p_c$ описывается функцией вида
\begin{equation}\label{eq:P_infty_beta}
P_\infty(p|p_c, \vb\beta) = \beta_1(p-p_c)^{\beta_2}.
\end{equation}
Тогда вектор коэффициентов $\vb\beta=(\beta_1, \beta_2)$ нетрудно определить из предельных соотношений \eqref{eq:lim_P_infty}.
В результате мощность перколяционного кластера при $p>p_c$ в первом приближении будет описываться функцией, степенной по доле достижимых узлов $p$ и показательно-степенной по порогу перколяции $p_c$:
\begin{equation}\label{eq:P_infty(p|pc)}
P_\infty(p|p_c) = \begin{cases}
\qquad 0, & p \leqslant p_c; \\
\left(\frac{p-p_c}{1-p_c}\right)^{1-p_c}, & p > p_c.
\end{cases}
\end{equation}

Так же как и при оценке порога перколяции на ограниченных решётках с непроницаемыми граничными условиями для оценки вероятности принадлежности произвольного узла к перколяционному кластеру $P_\infty$ можно использовать относительную частоту возникновения кластеров, стягивающих противолежащие границы решётки в заданном направлении.

\begin{figure}[tbh]\centering
\begin{overpic}[width=.3\w]{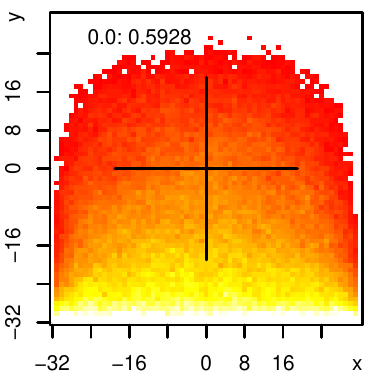}
\put(87,87){а)}
\end{overpic}\quad
\begin{overpic}[width=.3\w]{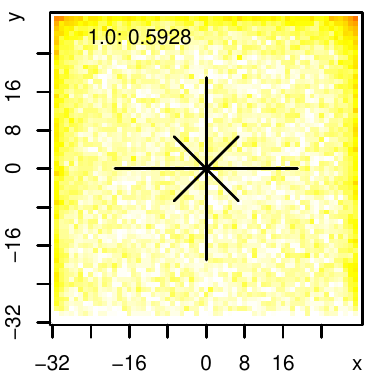}
\put(87,87){б)}
\end{overpic}\quad
\begin{overpic}[width=.3\w]{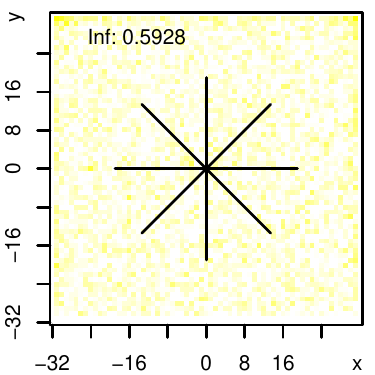}
\put(87,87){в)}
\end{overpic}\\
\begin{overpic}[width=.3\w]{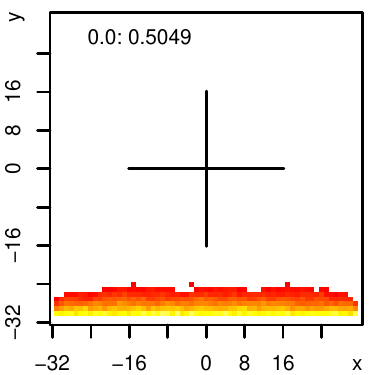}
\put(87,87){г)}
\end{overpic}\quad
\begin{overpic}[width=.3\w]{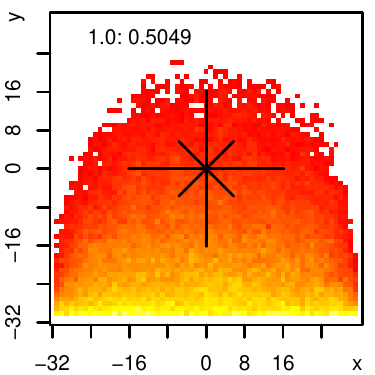}
\put(87,87){д)}
\end{overpic}\quad
\begin{overpic}[width=.3\w]{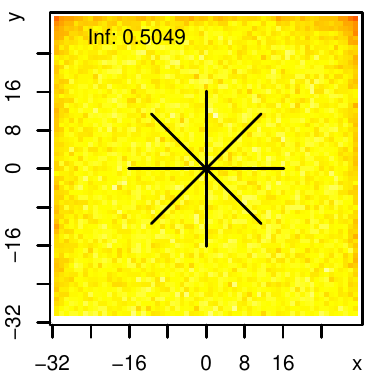}
\put(87,87){е)}
\end{overpic}\\
\begin{overpic}[width=.3\w]{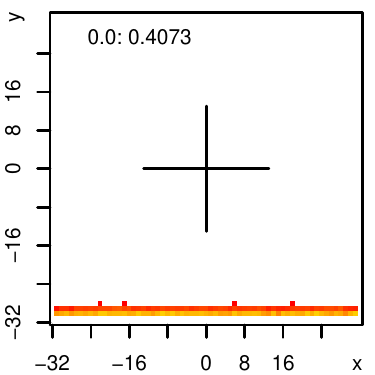}
\put(85,87){ж)}
\end{overpic}\quad
\begin{overpic}[width=.3\w]{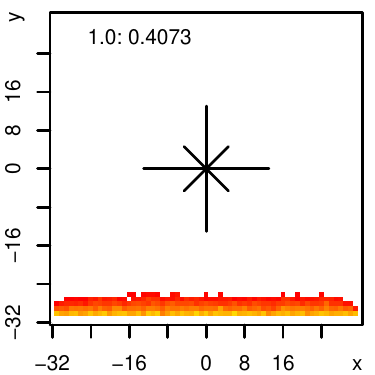}
\put(87,87){з)}
\end{overpic}\quad
\begin{overpic}[width=.3\w]{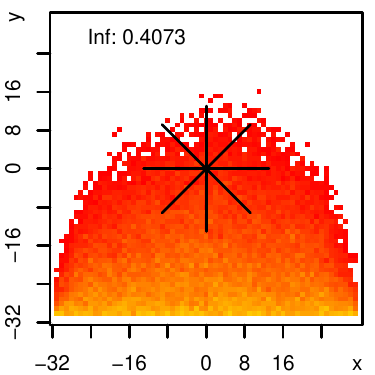}
\put(87,87){и)}
\end{overpic}
\caption{\label{pic:ssi2d_wxy}
Относительные частоты узлов $w_{xy}$ на двумерных квадратных решётках размером $L=65$ узлов с $(1, \pi)$-окрестностью при значениях доли достижимых узлов $p$ и показателя Минковского $\pi$: а-в)~$p=0,5928$; г-е)~$p=0,5049$; ж-и)~$p=0,4073$; а,г,ж)~$\pi=0$; б,д,з)~$\pi=1$; в,е,и)~$\pi\to\infty$}
\end{figure}

На рис.\,\ref{pic:ssi2d_wxy} показаны распределения относительных частот узлов двумерной квадратной перколяционной решётки размером $L=65$ узлов с $(1, \pi)$-окрестностью при различных значениях доли достижимых узлов $p$ и показателя Минковского $\pi$:  а-в)~$p=0,5928$; г-е)~$p=0,5049$; ж-и)~$p=0,4073$; а,г,ж)~$\pi=0$; б,д,з)~$\pi=1$; в,е,и)~$\pi\to\infty$. 
Использованные для построения реализаций стартовые подмножества состояли из достижимых узлов, расположенных вдоль нижней границы $y=-\frac{L}{2}$ этих решёток.
Объёмы выборок, используемых для расчёта относительных частот узлов $w_{xy}$ каждой решётки на рис.\,\ref{pic:ssi2d_wxy}, составляют $n=750$ реализаций.
Белый цвет соответствует узлам с относительной частотой $w_{xy}=0,6$, а красный~--- c относительной частотой $w_{xy}=0,2$.
Все узлы с относительными частотами, выходящими за указанный интервал, условно не показаны.

Нетрудно заметить, что распределения относительных частот $w_{xy}$, расположенные на рис.\,\ref{pic:ssi2d_wxy} вдоль главной диагонали, соответствуют критическим значениям доли достижимых узлов $p\approx p_c(\pi)$: 
а)~$p=0,5928\approx p_c(0)$; 
д)~$p=0,5049\approx p_c(1)$; 
и)~$p=0,4073\approx p_c(\infty)$.
В таком случае, распределения относительных частот $w_{xy}$, расположенные на рис.\,\ref{pic:ssi2d_wxy} ниже и выше главной диагонали, будут соответствовать до- и сверхкритическим значениям доли достижимых узлов $p\lessgtr p_c(\pi)$: б,в)~$p=0,5928>p_c(1)>p_c(\infty)$; г)~$p=0,5049<p_c(0)$; 
е)~$p=0,5049>p_c(\infty)$; ж,з)~$p=0,4073<p_c(0)<p_c(1)$.

Тогда оценкам мощности перколяционных кластеров $P_\infty$ в направлении оси $Oy$ будут соответствовать усреднённые относительные частоты подмножеств узлов, расположенных вдоль верхней $y=\frac{L}{2}$ границы перколяционных решёток.

\begin{figure}[tbh]\centering\small
\begin{overpic}{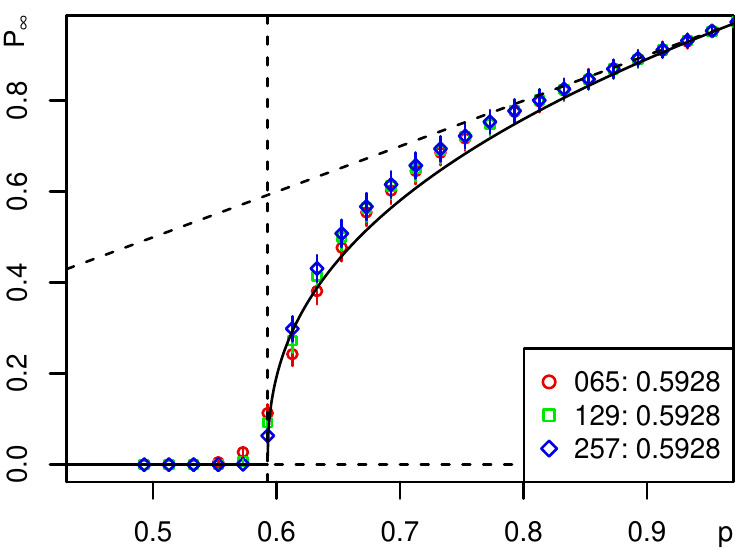}
\put(90,31){а)}
\end{overpic}\quad
\begin{overpic}{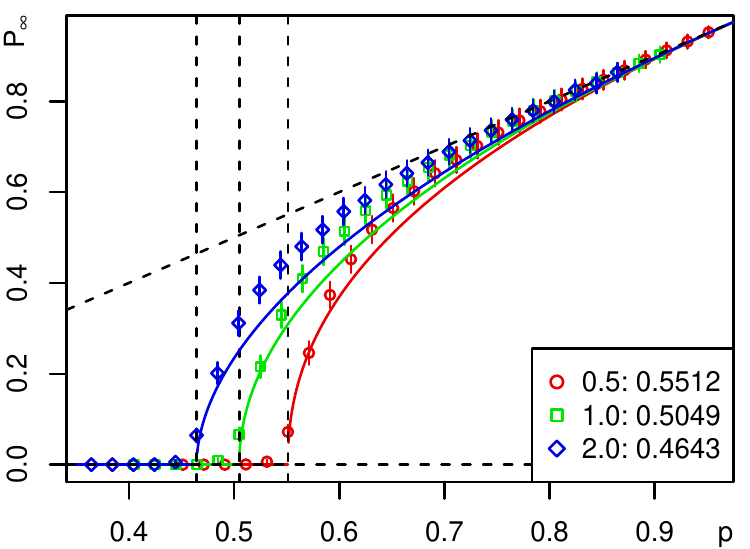}
\put(90,31){б)}
\end{overpic}
\caption{Мощности кластеров $P_\infty(p|L, \pi)$, стягивающих решётку от $y=-\frac{L}{2}$ до $y=\frac{L}{2}$ при: а)~$\pi=0$ и $L=65$, $129$, $257$ узлов; б)~$\pi=\frac{1}{2}$, $1$, $2$ и $L=129$ узлов}
\label{pic:ssi2d_pinf}
\end{figure}

На рис.\,\ref{pic:ssi2d_pinf} (а) приведены зависимости статистических оценок мощности перколяционных кластеров $P_\infty$ для двумерных квадратных решёток размерами $L=65$, $129$ и $257$ узлов с $(1, 0)$-окрестностью фон Неймана от доли достижимых узлов $p$.
Символами ``\scalebox{.8}{$\bigcirc$}'' показаны зависимости $P_\infty(p|L, \pi)$ для решётки размером $L_1\hm=65$ узлов, а символами ``$\square$'' и ``\raisebox{-.4ex}{\rotatebox{45}{$\square$}}''~--- зависимости $P_\infty(p|L, \pi)$ для решёток с размерами $L_2=129$ и $L_3=257$ узлов соответственно.

Объёмы выборок, используемых для расчёта мощности стягивающих кластеров $P_\infty$, для каждой тройки значений $(p|L, \pi)$ на рис.\,\ref{pic:ssi2d_pinf} составляет $n=750$ реализаций.
Точки соответствуют центрам, а вертикальные отрезки~--- $0,95$-доверительным интервалам \eqref{eq:Ci_w} для относительных частот $w$, усреднённых вдоль верхней границы решётки $y=\frac{L}{2}$.
Штриховые линии соответствуют прямым $p=p_c$, $P_\infty=0$ и $P_\infty=p$, а сплошные линии~--- функции \eqref{eq:P_infty(p|pc)} при $p_c=0,5928$.

На рис.\,\ref{pic:ssi2d_pinf} (б) приведены зависимости статистических оценок мощности перколяционных кластеров $P_\infty$ для двумерных квадратных решёток размером $L=129$ узлов с $(1, \frac{1}{2})$-, $(1, 1)$- и $(1, 2)$-окрестностью от доли достижимых узлов $p$.
Символами ``\scalebox{.8}{$\bigcirc$}'' на рис.\,\ref{pic:ssi2d_pinf} (б) показаны зависимости $P_\infty(p|L, \pi)$ для решётки с $(1, \frac{1}{2})$-окрестностью, а символами ``$\square$'' и ``\raisebox{-.4ex}{\rotatebox{45}{$\square$}}''~--- зависимости $P_\infty(p|L, \pi)$ для решёток с $(1, 1)$- и $(1, 2)$-окрестностями соответственно.

Сплошные линии соответствуют функции \eqref{eq:P_infty(p|pc)} где скейлинговые оценки порога перколяции $p_c(\pi)$ для двумерной квадратной решётки с $(1, \pi)$-окрестностью Мура были получены в предыдущем разделе: $p_c(\frac{1}{2})\approx 0,5512$, $p_c(1)\approx 0,5049$, $p_c(2)\approx 0,4643$.

\section{Заключение}

В настоящей работе представлены статистические методы построения оценок частоты $w$ и мощности перколяционных кластеров $P_\infty$, а также порога перколяции $p_c$ по выборочной совокупности реализаций на двумерных квадратных решётках с $(1, \pi)$-окрестностью Мура.

Порог перколяции $p_c$, наряду с размером решётки $L$ и долей достижимых узлов $p$, играет в вышеописанных моделях роль глобального управляющего параметра.
Все ранее построенные перколяционные модели допускали лишь дискретное управление по порогу протекания $p_c$ за счёт изменения топологии решётки.
Приведённые в настоящей работе оценки показывают, что изменение показателя Минковского $\pi$ обеспечивает непрерывное управление параметром $p_c(\pi)$ в достаточно широком диапазоне от $p_c(0)\approx 0,59282$ до $p_c(\infty)\approx 0,40728$, а линейно-логарифмическая зависимость для порога перколяции $p_c(\lb\pi)$ в первом приближении может быть аппроксимирована логистической функцией \eqref{eq:pc(lb_pi)}.

На рис.\,\ref{pic:ssi2d_pinf} хорошо заметно, что функция \eqref{eq:P_infty(p|pc)} при $p>p_c$ демонстрирует в основном качественное соответствие с результатами статистического моделирования.
Следуя работе \cite{moskaleff.info.2013.ssi20.pinf} можно предположить, что степенная аппроксимация \eqref{eq:P_infty_beta} без учёта топологических характеристик окрестности узла решётки не является достаточно адекватным средством и может быть использована для описания наблюдаемой зависимости мощности перколяционного кластера от доли достижимых узлов $P_\infty(p)$ в сверхкритической области лишь в первом приближении.
Действительно, евклидовы нормы векторов отклонений $\|e(L, \pi)\|_2$ для показанных на рис.\,\ref{pic:ssi2d_pinf} (а) статистических оценок $P_\infty$ от соответствующих значений аппроксимирующей функции \eqref{eq:P_infty(p|pc)} для двумерной квадратной решётки размером $L=65$, $129$ или $257$ узлов с $(1, 0)$-окрестностью фон Неймана имеют значения, вполне сопоставимые с оценками величины $P_\infty(p|L, \pi)$: $\|e(65, 0)\|_2=0,1605$; $\|e(129, 0)\|_2=0,1562$; $\|e(257, 0)\|_2=0,1532$.

В той же работе было отмечено изменение качества аппроксимации зависимости мощности изотропных перколяционных кластеров от доли достижимых узлов $P_\infty(p)$ с $(1, 0)$-окрестностью фон Неймана функцией \eqref{eq:P_infty(p|pc)} при переходе от двух- к трёхмерной решётке.
С формальной точки зрения это может объясняться двумя факторами: а)~увеличением числа узлов, образующих единичную окрестность фон Неймана, $n_3=6 > n_2=4$; б)~уменьшением критического значения доли достижимых узлов $p_{c3}\approx 0,3116 < 0,5927\approx p_{c2}$.
В таком случае, при моделировании зависимости мощности изотропных перколяционных кластеров $P_\infty$ от доли достижимых узлов $p$ с $(1, \pi)$-окрестностью Мура качество аппроксимации зависимости $P_\infty(p|\pi)$ функцией \eqref{eq:P_infty(p|pc)} также должно будет изменяться.

Сопоставление представленных выше результатов показывает, что качество аппроксимации эмпирической зависимости $P_\infty(p|\pi)$ функцией \eqref{eq:P_infty(p|pc)} с ростом показателя Минковского $\pi$ имеет тенденцию к понижению.
Действительно, евклидовы нормы векторов отклонений $\|e(L, \pi)\|_2$ для показанных на рис.\,\ref{pic:ssi2d_pinf} (б) статистических оценок $P_\infty$ от соответствующих значений аппроксимирующей функции \eqref{eq:P_infty(p|pc)} для двумерной квадратной решётки размером $L=129$ узлов с $(1, \pi)$-окрестностью Мура с ростом показателя Минковского $\pi$ также возрастают: $\|e(129, 0)\|_2=0,1562\hm< \|e(129, \frac{1}{2})\|_2\hm=0,1634\hm< \|e(129, 1)\|_2=0,1909< \|e(129, 2)\|_2\hm=0,2289$.

\end{document}